\documentclass[apj,numberedappendix,appendixfloats,tighten]{emulateapj}
\usepackage{graphicx}
\usepackage{amsmath}
\usepackage{appendix}

\newcommand{\ba}{\begin{array}}
\newcommand{\ea}{\end{array}}
\newcommand{\ben}{\begin{enumerate}}
\newcommand{\een}{\end{enumerate}}
\newcommand{\bei}{\begin{itemize}}
\newcommand{\eei}{\end{itemize}}

\newcommand{\f}[2]{\frac{#1}{#2}}

\newcommand{\denum}{(\Om-k_{\perp}\beta_{\perp})^{2}}
\newcommand{\denumx}{(\Om-k_{\perp}\beta_{\perp})^{-2}}

\newcommand{\DDir}{\relax{D\kern-.7em{/}}}

\newcommand{\ra}{\rightarrow}

\newcommand{\bB}{\textbf{B}}
\newcommand{\bE}{\textbf{E}}

\newcommand{\bk}{\mathbf{k}}

\newcommand{\be}{\begin{equation}}
\newcommand{\ee}{\end{equation}}
\newcommand{\bea}{\begin{equation*}}
\newcommand{\eea}{\end{equation*}}

\newcommand{\ave}[1]{\left\langle #1\right\rangle}

\newcommand{\pr}{\partial}

\newcommand{\nin}{\relax{\in\kern-.8em{/}}}

\newcommand{\te}{\theta}

\newcommand{\bt}{\beta}

\newcommand{\De}{\Delta}

\newcommand{\Om}{\Omega}
\newcommand{\om}{\omega}

\newcommand{\ep}{\epsilon}

\newcommand{\sref}{\S~\ref}

\newcommand{\oo}{\omega_0}
\newcommand{\ocr}{\omega_{\text{CR}}}

\newcommand{\jcr}[1]{J_{\text{CR}#1}}

\newcommand{\jo}[1]{J_{0#1}}
\newcommand{\Jo}{\mathbf{J}_0}
\newcommand{\gsh}{\Gamma_{\text{sh}}}
\newcommand{\etaz}{\eta_{0}}

\newcommand{\cre}{\epsilon}

\begin{document}
\title{Long wavelength unstable modes in the far upstream of relativistic collisionless shocks}
\author{Itay Rabinak, Boaz Katz and Eli Waxman}
\affiliation{Department of Particle Physics and Astrophysics, The Weizmann Institute of Science, Rehovot 76100, Israel}

\email{itay.rabinak@weizmann.ac.il}
\date{\today}

\begin{abstract}
The growth rate of long wavelength kinetic instabilities arising due to the interaction of a collimated beam of relativistic particles and a cold unmagnetized plasma are calculated in the ultra relativistic limit.
For sufficiently culminated beams, all long wave-length modes are shown to be Weibel-unstable, and a simple analytic expression for their growth rate is derived.
For large transverse velocity spreads, these modes become stable.
An analytic condition for stability is given.
These analytic results, which generalize earlier ones given in the literature, are shown to be in agreement with numerical solutions of the dispersion equation and with the results of novel PIC simulations in which the electro-magnetic fields are restricted to a given k-mode.
The results may describe the interaction of energetic cosmic rays, propagating into the far upstream of a relativistic collisionless shock, with a cold unmagnetized upstream.
The long wavelength modes considered may be efficient in deflecting particles and could be important for diffusive shock acceleration.
It is shown that while these modes grow in relativistic shocks propagating into electron-positron pair plasmas, they are damped in relativistic shocks propagating into electron-proton plasmas with moderate Lorenz factors $\gsh \lesssim (m_p/m_e)^{1/2}$.
If these modes dominate the deflection of energetic cosmic rays in electron-positron shocks, it is argued that particle acceleration is suppressed at shock frame energies that are larger than the downstream thermal energy by a factor of $\gtrsim \gsh$.

\end{abstract}
\maketitle

\section{Introduction}
\label{sec:introduction}

Current understanding of gamma-ray burst (GRB) "afterglows," the delayed low energy emission following the prompt $\gamma$-ray emission, suggests that the radiation observed is the synchrotron emission of energetic non-thermal electrons in the downstream of an ultra-relativistic collisionless shock driven into the surrounding interstellar medium (ISM) or stellar wind \citep{Zhang04,Piran04}.

This model requires a strong magnetic field and a large population of energetic electrons to be present in the downstream.
Observations suggest that the fraction of post-shock thermal energy density carried by non-thermal electrons, $\epsilon_e$, is large, $\epsilon_e\approx0.1$ \citep[e.g.][]{Zhang04,Frail01,Freedman01,Berger03}.
The fraction of post-shock thermal energy carried by the magnetic field, $\epsilon_B$, is less well constrained by observations.
However, in cases where $\epsilon_B$ can be reliably constrained by multi waveband spectra, values close to equipartition, $\epsilon_B\sim 0.01$ to $0.1$, are inferred \citep[e.g.][]{FWK00}.

The non-thermal energetic electron (and proton) population is believed to be produced by the diffusive (Fermi) shock acceleration (DSA) mechanism \citep[for reviews see][]{Drury83,Blandford87,Malkov01}.

The required magnetic fields in the shock frame in the downstream \citep[e.g.][]{FWK00} and upstream \citep{Zhuo06} regions are much larger than the ambient field, and thus require substantial
amplification. The accelerated particles are likely to have an important role in generating and maintaining the inferred magnetic fields.

The main challenge associated with the downstream magnetic field is that the field amplitude must remain close to equipartition deep into the downstream, over distances $\sim10^{10}l_{sd}$ \citep{Gruzinov99,Gruzinov01a}.
While near equipartition fields on skin depth scale are likely to be produced in the vicinity of the shock by electromagnetic (e.g. Weibel-like) instabilities \citep[e.g.][]{Blandford87,Gruzinov99,Medvedev99,Wiersma04}, they are expected to decay within a few skin-depths downstream \citep{Gruzinov01a}.
This suggests that the correlation length of the magnetic field far downstream and possibly upstream must be much larger than the skin depth, $L\gg l_{sd}$, perhaps even of the order of the distance from the shock \citep{Gruzinov99,Gruzinov01a,Katz07}.

The search for a self-consistent theory of collisionless shocks has led to extensive numerical studies using the particle in cell (PIC) based algorithms
\citep[e.g.][]{Gruzinov01a,Gruzinov01b,Medvedev05,Silva03,Nishikawa03,Frederiksen04,Jaroschek04,Spitkovsky05,Spitkovsky08,Martins09}. Such simulations have provided compelling evidence for acceleration of particles and generation of long lasting near-equipartition magnetic fields.
However numerically simulating the long term behavior is challenging and is currently restricted to pair ($e^+e^-$) plasmas in 2D \citep[e.g.][]{Spitkovsky08b,Keshet09}.

Large scale magnetic fields may possibly be generated in the upstream by the interaction of the beam of CRs propagating ahead of the shock and the upstream plasma \citep[e.g.][]{Katz07,Keshet09}.
In particular high energy CRs naturally introduce large scales due to their large Larmor radius, and the large distances to which they propagate into the upstream.
Instabilities arising from the interaction of relativistic beams and cold plasmas have long been studied
\citep[][and references therein]{Akhiezer75} and are suspected of amplifying the magnetic field in the shock transition layer
\citep{Gruzinov99,Medvedev99,Wiersma04,Bret05,Lyubarsky06,Achterberg07,Achterberg07b,Bret09}.
In \citep{Lemoine09} a systematic study of these instabilities for the lowest energy CRs, with energies comparable to the thermal energy of the shocked plasma, and their application for Fermi acceleration is given.

In this paper we analyze long wavelength plasma instabilities resulting from the counter-streaming flow of high energy CRs ,$\gamma \gg \gsh$, running far ahead of the shock and a non-magnetized upstream plasma.
The analysis is restricted to long wavelength modes, $ k \ll \oo / c$, which are expected to deflect particles
efficiently.
For simplicity it is assumed that the particle distribution is homogenous.
The paper is organized as follows.
In \sref{sec:analysis} we calculate the growth rate of long wavelength modes.
We separately discuss highly collimated beams and beams with a significant transverse velocity spread, and derive a condition for the stability of these modes.
In \sref{sec:App2CollSh} we discuss the possible implications of these results to collisionless shocks.
In \sref{sec:discussion} we summarize the main results and conclusions. An estimate of the saturation level of the modes is beyond the scope of this paper.
Throughout this paper, units with $c=1$ are assumed ($c$ is retained in some of the expressions).

\section{Analysis}
\label{sec:analysis}

Consider a homogenous, anisotropic distribution of particles consisting of a cold plasma and an axi-symmetric beam of ultra relativistic CRs.
The analysis is carried out in the rest frame of the cold plasma which initially has zero magnetic and electric fields.

The plasma frequencies, in this frame, of the cold plasma and the beam are denoted by $\oo$ and $\ocr$, respectively, where the plasma frequency of a plasma with species $i$ is defined by
\begin{equation} \label{eq:ompDef}
\om_p^2 = \sum_i  \frac{4 \pi q_i^2}{m_i} \int \frac{\text{d}^3 p'}{\gamma(p')} f_i(p'),
\end{equation}
where $q_i, m_i, f_i(p) $ are the species' charge, mass and momentum distribution.
It is assumed that $\oo \gg \ocr$, and that the CRs have a small but finite spread in the velocity directions.

In this section we analyze the linear growth of unstable modes with long wavelengths, $k \ll \oo$.
We start by considering a beam with no transverse velocity spread in \sref{sec:nospread}.
We show that the entire k-space regime considered is unstable and provide a simple analytic expression of the instability growth rate.
The effects of a spread in the velocity directions of the CRs are discussed in \sref{sec:spread}.

\subsection{No spread}\label{sec:nospread}

Consider the simplest case in which all the particles in the beam propagate in the same direction and are ultra relativistic (with an arbitrary energy distribution).
It is straight forward to write the full dispersion equation which turns out to be a sixth order polynomial equation for $\om$ with real coefficients
\citep[cf. \sref{sec:DispEqSol} and e.g.][\S~6.4.2]{Akhiezer75}.
Four of the six solutions for $\om$ are small perturbations, of order $\ocr^2/\oo^2$, of the four cold plasma oscillating modes $\om=\pm \oo, \pm (\oo^2 + k^2)^{0.5}$, and are real (stable).
The two remaining solutions have a non zero imaginary part and therefore  are complex conjugates of each other.
Hence for each $\bold{k}$ there is one unstable mode.
Below we derive, directly from the Maxwell-Vlasov equations, an approximate expression for the growth rate of this mode, Eqs. \eqref{eq:GR}, and \eqref{eq:OmReal}.
More details and a numerical solution for the dispersion equation are given in \sref{sec:DispEqSol}.

For any axi-symmetric distribution of particles the linear modes can be separated into modes having an electric field in the $x-k$ plane, where $x$ is the axis of symmetry (and magnetic field perpendicular to this plane), and modes with an electric field perpendicular to $x$ and $k$ (cf. \sref{sec:DispEqSol}).
The unstable mode has an electric field, $\bE$, in the $x-k$ plane, and a magnetic field, $B$, perpendicular to this plane.
The electrical currents carried by the cold plasma and the CRs,  as derived by the Vlasov equations, are respectively given by
\be
    \label{eq:J0N}
    4\pi \Jo = \frac{\oo^{2}}{-i\omega}\mathbf{E},
\ee
and
\begin{align}
    \label{eq:Jcr_perpN}
    4\pi \jcr{,\perp}   &= i\ocr^{2}(B-E_{\perp})/\Omega; \\
    \label{eq:Jcr_parN}
    4\pi \jcr{,||}      &=-i\ocr^{2}(B-E_{\perp}) \, k_{\perp}/\Omega^{2},
\end{align}
where subscripts ${\perp}$ and ${||}$ correspond to components that are perpendicular to the beam and parallel to the beam respectively, $\Omega\equiv k_{||}-\omega$, and where we used the ultra relativistic approximation $\beta = 1$ for the CRs.

By neglecting the displacement current $\pr_t E$, compared to the current carried by the cold plasma [using $\om \ll \oo$ which is self consistently implied by the result, Eqs. \eqref{eq:GR} and \eqref{eq:OmReal}], the Maxwell equations read
\begin{align}
    \label{eq:Amper_perpN}
    i k_{\perp} B &= 4\pi(\jcr{,||}+\jo{,||}); \\
    \label{eq:Amper_parN}
    -ik_{||}B   &= 4\pi(\jcr{,\perp}+\jo{,\perp});\\
    \label{eq:FaradayN}
    i\omega B &= -ik_{\perp}E_{||}+ik_{||}E_{\perp}.
\end{align}
Equation \eqref{eq:FaradayN} can be written as
\begin{equation} \label{eq:FaradayN1}
    i \omega (B - E_{\perp}) = -ik_{\perp}E_{||} ,
\end{equation}
by self consistently neglecting $ \Om E_{\perp}$ as follows.
In regimes where $\Omega  \ll k_{||}$, this term is negligible compared to the term $k_{||}E_{\perp}$. Otherwise, where $\Omega \gtrsim k_{||}$, Eq. \eqref{eq:OmReal}, implies that $\Om\sim\om$ and Eqs. \eqref{eq:Amper_parN} and \eqref{eq:Jcr_perpN} imply that $E_{\perp}\ll B$, making the term $ \Om E_{\perp}$ negligible compared to $\om B$.

By neglecting the term $i k_{\perp} B$ in Eq. \eqref{eq:Amper_perpN}, compared to $ \jcr{,||} $ [using Eqs. \eqref{eq:Jcr_parN} and \eqref{eq:GR} ], and substituting for $\jcr{,||}, \jo{,||}, E_{||}, $  equations \eqref{eq:Jcr_parN}, \eqref{eq:J0N}, and \eqref{eq:FaradayN1} respectively, equation \eqref{eq:Amper_perpN} becomes
\be
    \label{eq:Amper_PerpNSub}
    \ocr^{2}B \, k_{\perp}/\Omega^{2} =
    -\oo^{2} B / k_{\perp}.
\ee
The solution of this equation for $\om$ is
\begin{align}
    \label{eq:GR}
    Im\{\omega\} &= \pm\frac{\ocr}{\oo} k_{\perp} \equiv \pm\eta_0(k_{\perp});\\
    \label{eq:OmReal}
    Re\{\omega\} &= k_{||}.
\end{align}
Eq. \eqref{eq:Amper_parN} determines the ratio between $E_{\perp}$ and $B$.
The result, Eqs. \eqref{eq:GR} and \eqref{eq:OmReal}, agrees with that of \citep[][\S~6.4.2]{Akhiezer75} and \citep{Lemoine09} in the relevant $k$ space regimes.

By retaining the term $i k_{\perp } B$ in Eq. \eqref{eq:Amper_perpN}, and assuming finite Lorentz factors, the dispersion relation can be similarly solved resulting in
\begin{equation} \label{eq:GW1}
    Im\{\omega\} = \pm
    \ocr \sqrt{
    \frac{k_{\perp}^2}{\oo^2 + k_{\perp}^2}  +
    k_{||}^2 \ave{\frac{1}{\gamma^2}} },
\end{equation}
where averaging is over $q_i^2 f_i(p) / (m_i \gamma) $ [cf. Eq. \eqref{eq:ompDef}].
This solution, which is valid for all $k$  with $k_{||} \ll 1$,  reduces to Eq. \eqref{eq:GR} in the regime
\begin{equation}
    \label{eq:GWRange}
    k_{||}\ave{\gamma^{-2}}^{1/2} \ll k_{\perp} \ll \oo.
\end{equation}

\begin{figure}
\includegraphics[scale=0.8]{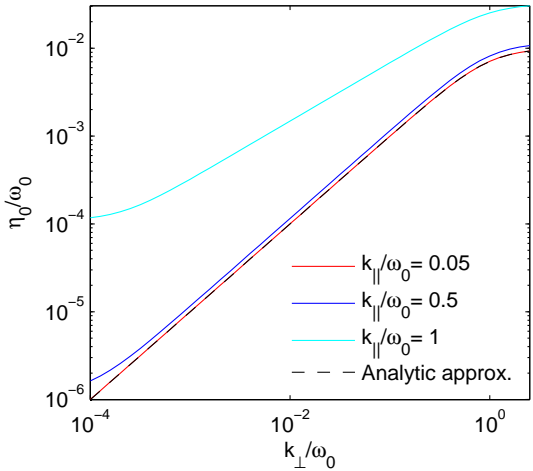}
\caption{\label{fig:GrDeltaFunc}
Growth rate of unstable modes for a delta function momentum distribution of beam particles with plasma frequency $\ocr / \oo = 0.01$ and Lorentz factor $\gamma = 5000$.
The solid lines are the exact solutions of the dispersion relation, evaluated numerically (cf. \sref{sec:DispEqSol}.)
The dashed line (covering the solid line with $k_{||}=0.05\oo$) is the analytical approximation given in equation \eqref{eq:GR}.
The curvature at low values of $k_{\perp}$ is due to the electrostatic
mode.}
\end{figure}

In figure \eqref{fig:GrDeltaFunc} the analytical approximation for the growth rate
given in equation \eqref{eq:GR} is compared with a numerical solution of equation \eqref{eq:FullDisp} for different modes.
For illustration modes with $k_{||} \sim \oo$ that are not analyzed in this paper are shown.
As can be seen in the figure, the analytic expression provides an excellent approximation in the relevant k-space regime.

\subsection{With spread} \label{sec:spread}
For a velocity distribution of the beam particles that differs from a delta function,
Eq. \eqref{eq:Amper_PerpNSub} should be replaced with
\be
    ik_{\perp}B = -i\ocr^{2}B \, k_{\perp} \left\langle \frac{1}{\denum}\right\rangle
    -i \oo^{2} B / k_{\perp},
\ee
or
\be \label{eq:DispSpread}
    \left\langle \frac{1}{(\Om  -k_{\perp}\beta_{\perp})^{2}}\right\rangle = -\frac{1}{\etaz^2},
\ee
where the same approximations leading to Eq. \eqref{eq:Amper_PerpNSub} were used, $\Omega\equiv k_{||}\beta_{||}-\omega$, and $\etaz$ is the growth rate for a delta function momentum distribution Eq. \eqref{eq:GR}.
Averaging is carried over the velocity distribution function,
\be
    \label{eq:MeanOm}
    \left\langle \frac{1}{\denum}\right\rangle = \frac{1}{\ocr^2}
    \int_{-\infty}^{\infty}\frac{f(\beta_{\perp})}{\denum}d\beta_{\perp},
\ee
where
\begin{equation}
    f(\beta_{\perp}) = \sum_i \frac{4 \pi q_i^2}{m_i}
    \int \frac{\text{d}^3 p'}{\gamma(p')} f_i(p')
    \delta(\beta_{\perp}' - \beta_{\perp}).
\end{equation}
It is to be understood that whenever a singularity is encountered, the expression should be evaluated at $\Om  \rightarrow \Om  -i \epsilon$  in the limit $\epsilon \rightarrow  0^+ $.

Consider first the following 1D rectangular distribution: $\beta_{x}\equiv\beta_{||}=\beta\cos(\theta),$
$\beta_{y}\equiv\beta_{\perp}=\beta\sin(\theta),$ $\beta_{z}=0,$
with $\theta$ uniformly distributed between $\pm\Delta\theta$, and
$\beta=\sqrt{1-\gamma^{-2}}$
\citep[see][for a discussion of a distribution with two identical counter streaming beams]{Achterberg07}.
For small angular spread the growth rate can be approximated analytically by neglecting the variations of $\beta_{x}$ (assuming $\beta_{x}=\beta$).
Under this assumption Eq. \eqref{eq:MeanOm} reads
\be \left\langle \frac{1}{\denum}\right\rangle=
\frac{1}{(k_{\perp}\beta\Delta\theta)^{2}-\Om^{2}},
\ee

the dispersion relation [Eq. \eqref{eq:DispSpread}] reads \begin{equation}
\frac{1}{(k_{\perp}\beta\Delta\theta)^{2}-\Om^{2}}=-\frac{1}{\etaz^2},\end{equation}
and the solution for $\Om^{2}$ is
\begin{equation}\label{eq:1DspreadSolution}
\Om^{2}=(k_{\perp}\beta\Delta\theta)^{2}-\etaz^2.
\end{equation}

\begin{figure}
\includegraphics[scale=1]{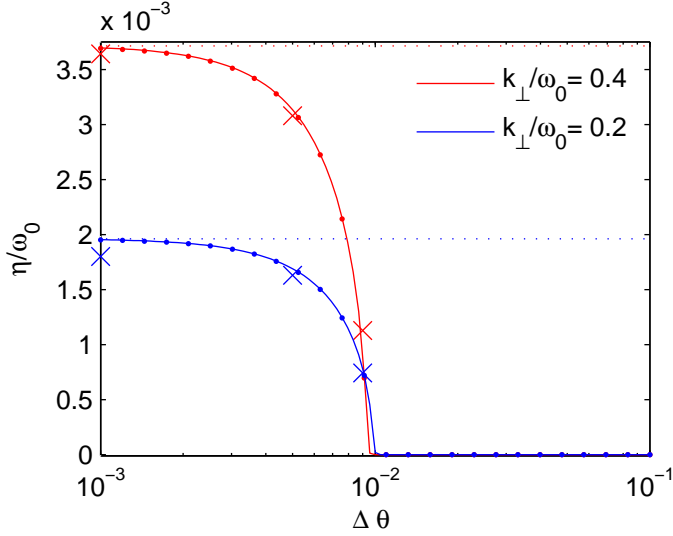}
\caption{\label{fig:GRwithSpread}
Growth rate calculations in the presence of a spread in the transverse velocities of the beam particles.
Different methods, as described in the text in \sref{sec:spread}, were used to calculate the growth rates.
Results are given for  $\{k_{||}/\oo=0.05, k_{\perp}/\oo = 0.4 (0.2)\}$ Weibel modes in red (blue) of a beam with $\gamma_u = 5000$, $\ocr/\oo = 10^{-2}$ and different spreads, $\Delta \theta$.
Solid lines, dots and x symbols give the growth rates obtained using the semi-analytical calculation, Eq. \eqref{eq:1DspreadSolution}, and a '1 mode PIC simulation' (cf. \sref{sec:NumMethods}) respectively.
Dotted lines show the growth rate for a delta function momentum distribution (no spread, $\Delta \theta=0$).
}
\end{figure}

For illustration, the growth rates of the unstable modes are shown in figure \eqref{fig:GRwithSpread} as a function of the spread $\Delta \theta $ for the 1D rectangular distribution considered above.
The solid lines in this figure are the results of a semi-analytical calculation in which the velocity integrals where evaluated analytically, and a continuous solution for the dispersion equation as a function of the spread was found numerically, starting from the solution for a delta function distribution.
The obtained solution is verified to be the fastest growing one at a given $\bold{k}$, by comparing it to the results of a '1-mode' PIC simulation (shown as x signs, cf. \sref{sec:NumMethods}).
These results are compared to the growth rate estimate of Eq. \eqref{eq:1DspreadSolution} (dots) and to the growth rate of the delta function $\eta_0$ distribution (dotted line).
As can be seen in the figure the estimate given in Eq. \eqref{eq:1DspreadSolution} is in very good agreement with both the simulation and the semi-analytical calculation.

Note the following features of the solution given by Eq. \eqref{eq:1DspreadSolution}:
\begin{enumerate}
\item For a sufficiently small spread, $\Delta \theta<\Delta\theta_{\text{crit}}$ there exists an unstable mode with $Re\{\omega\}= k_{||}\bt$, where the condition for instability is
    \begin{equation} \label{eq:StabilCriter1D}
    k_{\perp} \bt \Delta\theta < \etaz,
    \end{equation}
    or
     \begin{equation} \label{eq:StabilCriter1Dom}
    \bt \Delta\theta < \ocr/\oo
    \end{equation}
\item The growth rate, $Im(\Om)$ monotonically decreases from $\etaz$ to 0 as the spread $\Delta \theta$ is increased.
\item For $\Delta \theta>\Delta\theta_{\text{crit}}$ the mode becomes stable.
\end{enumerate}

The criterion for instability given in \eqref{eq:StabilCriter1D} can be interpreted as the simple requirement \citep[][\S~6.4.2]{Akhiezer75} that the particles in the beam do not move in the direction perpendicular to the beam a distance exceeding the wavelength of the mode during one e-folding time (of the delta-function instability).

We next show that most of the features of equation \eqref{eq:1DspreadSolution} are generic to a large class of axi-symmetric distributions of CRs with a small spread.

For purely imaginary $\Om=-i\eta$, Eq. \eqref{eq:MeanOm} reads
\begin{equation}
    \left\langle \frac{1}{\denum}\right\rangle=
    2\int_{0}^{\infty}\frac{k_{\perp}^{2}\beta_{\perp}^{2}-\eta^{2}}
    {(\eta^{2}+k_{\perp}^{2}\beta_{\perp}^{2})^{2}}
    f(\beta_{\perp})d\beta_{\perp},
\end{equation}
and has a zero imaginary part for all real $\eta$.
In the limits $\eta\ra 0^+, \infty $ the term $\ave{\denumx}$ goes to
$-k_{\perp}^{-2}\ave{{\beta_{\perp}^{-2}}} $  and 0 respectively, where
\begin{align}\label{eq:btavdeff}
    \ave{{\f{1}{\beta_{\perp}^{2}}}} & \equiv
    -\lim_{\ep \ra 0}
    \int_{-\infty}^{\infty}\frac{f(\beta_{\perp})}{(\beta_{\perp}+ i\ep) ^{2}}d\beta_{\perp} \\
    & = \int_{-\infty}^{\infty}
    \frac{f(\beta_{\perp})|_{\beta_{\perp} =0} - f(\beta_{\perp}) }{\beta_{\perp}^{2}}d\beta_{\perp}.
\end{align}
The last equality follows from the fact that
\be
    \int_{-\infty}^{\infty} \f{1}{(x + i\ep)^2} dx = 0.
\ee

The following features of Eq. \eqref{eq:DispSpread} generalize the features of the 1D distribution considered above with the condition for instability,
\begin{equation}\label{eq:StabilCriter}
    k_{\perp}\ave{\beta_{\perp}^{-2}}^{-1/2} < \etaz,
\end{equation}
or
\begin{equation}\label{eq:StabilCriterom}
    \ave{\beta_{\perp}^{-2}}^{-1/2} < \ocr/\oo,
\end{equation}
generalizing the condition \eqref{eq:StabilCriter1D}.
\begin{enumerate}
\item For a small spread, $\ave{\beta_{\perp}^{-2}}^{-1/2} < \etaz/k_{\perp}$ (equivalent to $k_{\perp}^{-2}\ave{{\beta_{\perp}^{-2}}} > 1 / \etaz^2$), there exists an unstable mode, with $Re\{\omega\}= k_{||}\bt$. To see this, note that the term $\ave{\denumx}$ continuously changes from $-k_{\perp}^2\ave{\beta^{-2}}$ to  0 as $\eta$ changes from 0 to $\infty$, and must be equal to $-\etaz^2$ for some positive $\eta$.
\item Assuming that the term $\ave{\denumx}$ is monotonically increasing with $\eta$, for a marginal spread $\ave{\beta_{\perp}^{-2}}^{-1/2} \ra \etaz/k_{\perp}$ the mode becomes stable $\eta \ra 0 $. This suggests that for larger spreads the mode is stable.
\end{enumerate}

As for the 1-D distribution considered above, the criterion for instability given in \eqref{eq:StabilCriter} can roughly be interpreted as
the requirement that the particles in the beam do not move in the direction perpendicular to the beam a distance exceeding the wavelength of the mode during one e-folding time. Note however that the simple expression, $\De \te
\bt$, in the 1-D case, which is equal to the maximal velocity of the particles in the direction perpendicular to the beam, is replaced by the non trivial average, $\ave{\beta_{\perp}^{-2}}^{-1/2}$ [the velocity average defined
in \eqref{eq:btavdeff}], which has a less obvious meaning.

\section{Application to collisionless shocks}
\label{sec:App2CollSh}

We next discuss the possible application of the above results to the study of long wavelength magnetic field generation in the far upstream of collisionless shocks.
The application of linear analysis of homogenous distributions to the non-homogenous and non-linear problem of collisionless shocks is far from being trivial.
Furthermore, the long wavelength modes studied above are not the fastest growing modes, and can be affected by the faster growing, short wavelength modes once the latter reach the non-linear stages.
Nevertheless, the analysis of linear growth of long-wave length modes is an important step in the study of long-wave length magnetic field generation and can be used as a basis for comparison once more accurate calculations are made (e.g. PIC simulations).
In addition it is possible that some of the main features of the linear modes also appear in the more complicated shock scenario.

Consider a shock with Lorentz factor $\gsh \gg 1$ propagating into a cold plasma with particle density $n_0$ and plasma frequency $\oo$.
We first assume that all particles in the plasma have the same mass (e.g. electrons-positrons) and then generalize the results to an electron proton plasma.

Assume that high energy, shock accelerated cosmic rays carry a fraction $\sim\epsilon_{p}$ of the post-shock energy. In the shock frame, the cosmic rays are not highly beamed and have an energy distribution
$ n_{cr,s} (>\gamma_{s})\sim\epsilon_{p} \gsh n_{0}(\gamma_{s}/\Gamma_{sh})^{-p+1} $
with $p\approx2$, where $\gamma_{s}$ is the Lorenz factor of the cosmic rays in the shock frame.

We wish to analyze the generation of long wavelength magnetic fields in a region surrounding a point, $x$, deep in the upstream due to the interaction of the CRs that reach this point and the incoming upstream particles.
We assume that this point is reached by a substantial fraction of the CRs that have Lorenz factors larger than a space dependent minimum  $\gamma_{s}(x)$, and study the instabilities in a simplified homogenous model of the upstream frame.

The cosmic rays in the upstream frame are beamed to an angular separation of $\sim 1/\Gamma_{sh}$, and have a plasma frequency
\begin{equation}\label{eq:OmCr}
    \ocr \sim \epsilon_{p}^{1/2}(\gamma_{s}/\Gamma_{sh})^{-p/2} \omega_{0},
\end{equation}
where $\oo$ is the upstream plasma frequency and different values of $\gamma_{s}$ represent different positions in the simplified picture\footnote{Note that the energy carried by the cosmic rays in this frame greatly exceeds that of the upstream particle rest mass energy density  for $\Gamma_{s} \gg \epsilon_{p}^{1/4}$, $e_0=n_0mc^2$.}.
Deep in the upstream, where $\gamma_s(x) \gg \gsh$, the plasma frequency of the cosmic rays is much smaller than that of the upstream, and the analysis of section \sref{sec:analysis} holds.

Eqs. \eqref{eq:GR} and \eqref{eq:OmCr} imply that long wavelength modes will grow with a growth rate of approximately
\be\label{eq:CollshockGR}
    \eta_0=k_{\perp} \epsilon_{p}^{1/2} (\gamma_{s}/\Gamma_{sh})^{-p/2}.
\ee

It is useful to compare the e-folding time $\etaz^{-1}$ of this instability to the time it takes the ambient magnetic field $B_0$ to deflect a cosmic ray particle by an angle of $1/\gsh$ back to the
downstream, ${T_{R,0}\sim \gsh\gamma_s mc/(eB_0\gsh)=\gamma_s mc/(eB_0)}$.
The ratio of the two times is
\begin{equation} \label{eq:etazt}
     T_{R,u}\etaz \sim \gsh \frac{k_{\perp}}{\oo}
    \epsilon_{p}^{1/2} \epsilon_{B,0}^{-1/2}
    (\gamma_{s}/\Gamma_{sh})^{1-p/2},
\end{equation}
where $\epsilon_{B,0}\sim B_0^2/(8\pi n_0 m c^2)$ is the ratio of the energy density in the ambient magnetic field to the upstream rest mass energy
density. For small values of the magnetic field in the upstream, there will be a large range of $k$ vectors
\begin{equation} \label{eq:ketaT}
k_{\perp}\gtrsim \gsh^{-1}\epsilon_{p}^{-1/2}\epsilon_{B,0}^{1/2}
\end{equation}
for which the instability will grow on time scales
that are much shorter than the deflection time of the particles.

Far away from the shock in the upstream, where $\ocr$ is sufficiently small, the instability will be suppressed due to the $1/\Gamma_{sh}$ spread in the cosmic rays transverse velocities.
Using Eqs. \eqref{eq:StabilCriter} [or \eqref{eq:StabilCriter1D}] and \eqref{eq:CollshockGR}, the modes are unstable only at locations in the upstream
where
\begin{equation}\label{eq:SimpInstabCond}
\gamma_{s}(x) \lesssim \Gamma_{sh} \left(\epsilon_{p}^{1/2}\Gamma_{sh}
\right)^{2/p}.
\end{equation}

\begin{figure}
\includegraphics[scale=1]{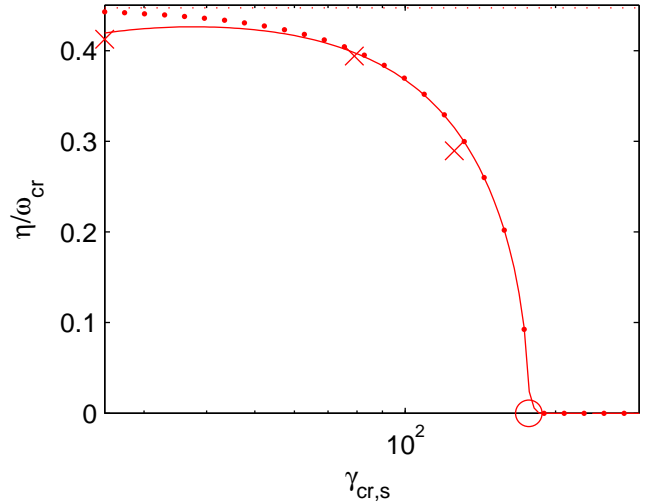}
\caption{ \label{fig:GRforRealBeam}
Growth rate for a specific Weibel mode, with $k_{||}/\oo=0.05$ and $ k_{\perp}/\oo = 0.5$, shown as a function of the shock frame Lorentz factor, $\gamma_{s}$, of the beam particles.
The beam particles have a 1D rectangular phase space distribution (cf. \sref{sec:spread}) with parameters as discussed in the text in \sref{sec:App2CollSh}.
The solid line, dots and x symbols give the growth rates obtained using the semi-analytical calculation (cf. \sref{sec:spread}), Eq. \eqref{eq:1DspreadSolution}, and a '1 mode PIC simulation' (cf. \sref{sec:NumMethods}) respectively.
The dotted line is the growth rate obtained for a delta function distribution, $\eta_0$.
The red circle shows the value of $\gamma_{\rm CR,s}$ given by equation \eqref{eq:SimpInstabCond}, beyond which the modes are predicted to be stable.
}
\end{figure}

For illustration, the growth rate of a specific unstable mode is shown in figure \eqref{fig:GRforRealBeam} as a function of $\gamma_{s}$.
For simplicity, the momentum distribution is assumed to be the same as in \sref{sec:spread} with parameters chosen in terms of shock parameters as: $\pm\Delta\theta=\pm1/\Gamma_{sh}$ (with $\gsh = 25$), upstream Lorentz factor $\gamma_{u} = 2\gamma_{s}\Gamma_{sh}$, and beam plasma frequency, $\omega_{cr}$ as given in Eq. \eqref{eq:OmCr} with $p=2$, and $\epsilon_p =0.1$.
In the figure we also show the results of the semi-analytical calculation (cf. \sref{sec:spread}), the growth rate estimate of Eq. \eqref{eq:1DspreadSolution} (dots) and the results of the 1-mode PIC simulation (cf. \sref{sec:NumMethods}, x signs).
For comparison the growth rate $\eta_0$ of the delta function distribution (cf. \sref{sec:nospread}, dotted line), and the maximal $\gamma_s$ from equation
\eqref{eq:SimpInstabCond} (red circle, y axis value arbitrarily set to 0) are also shown. As can be seen in the figure the delta function momentum distribution result is a good approximation at small angular spread, and the angular spread suppresses the growth rate at high cosmic ray Lorentz factors in accordance with the estimate in equation \eqref{eq:SimpInstabCond}.

Next, consider a shock propagating into a electron-proton plasma.
The equation for the plasma frequency of the beam \eqref{eq:OmCr} should be replaced by
\begin{equation}\label{eq:OmCrp}
    \ocr \sim \epsilon_{p}^{1/2}(m_e/m_p)^{1/2}(\gamma_{s,p}/\Gamma_{sh})^{-p/2} \omega_{0},
\end{equation}
where $\gamma_{s,p}$ is the minimal CR proton shock frame Lorentz factor at the position considered and where the contribution of the electron CRs at a given particle energy was neglected.
Given the transverse velocity spread, $1/\Gamma_{sh}$, of the CRs Eqs. \eqref{eq:StabilCriterom} and \eqref{eq:StabilCriter1Dom} imply that the long wavelength modes do not grow for moderate shock Lorentz factors \citep[see also][]{Lyubarsky06}
\begin{equation}\label{eq:MinimalGepr}
    \gsh \lesssim 100 \epsilon_{p,-1}^{1/2} ,
\end{equation}
where $\epsilon_{p} = 0.1 \epsilon_{p,-1}.$
For higher shock Lorentz factors, the long wavelength modes grow at locations where the minimal shock frame CR energy, $\cre_{s}(x)$ (equal for CR electrons and protons), is sufficiently low [cf. Eq. \eqref{eq:SimpInstabCond}]
\begin{equation}\label{eq:EPInstabCond}
    \cre_{s}(x) \lesssim \Gamma_{sh} m_p
    \left(\epsilon_{p}^{1/2}  \Gamma_{sh}
    \right)^{2/p} (m_e/m_p)^{1/p}.
\end{equation}

\section{Discussion}
\label{sec:discussion}

In this paper the growth rates of the long wavelength unstable modes arising from the interaction of a beam of ultra-relativistic CRs and an unmagnetized cold plasma were calculated.
We have shown that in the ultra-relativistic limit all long wavelength modes are unstable with a growth rate $\etaz = k_{\perp} \ocr / \oo $ [Eq. \eqref{eq:GR}], as long as the spread in the transverse velocity distribution of the beam is sufficiently small.
An extension of this result, for finite Lorentz factors and $k_{\perp} \gtrsim \oo  $, is given in Eq. \eqref{eq:GW1}.
For large transverse velocity spreads the instability is suppressed.
The condition for instability was derived for a large class of velocity distributions [Eqs. \eqref{eq:StabilCriterom} and \eqref{eq:btavdeff}].

The possible application of these results to the interaction of CRs with the incoming plasma in the far upstream of unmagnetized collisionless shocks was addressed in \sref{sec:App2CollSh}.
In a shock propagating into an electron positron plasma, the instability grows in upstream regions where the minimal shock frame CR Lorentz factor is sufficiently low, $\gamma_{s}<\gsh^2 \ep_p^{1/2}$ [cf. Eq. \eqref{eq:SimpInstabCond}].
Farther upstream of the shock the instability is suppressed due to the low density of the CRs.
If these instabilities dominate the particle deflections responsible for particle acceleration, the acceleration of CRs to shock frame Lorentz factors exceeding $\gsh^2 \ep_p^{1/2}$ [cf. Eq. \eqref{eq:SimpInstabCond}] may be obstructed.

The long wavelength instabilities considered do not grow in the upstream of a shock propagating into an electron proton plasma having Lorentz factors $\lesssim 100$ \citep[cf. Eq. \eqref{eq:MinimalGepr}, see also][]{Lyubarsky06} implying that:
1. The modes considered are not important in the relativistic shocks responsible for the observed GRBs afterglow emission (except possibly at the earliest stages, when $\gsh \gtrsim100$); 2. Particle acceleration in electron-positron plasmas suggested by the results of PIC simulations \cite{Spitkovsky08b}, where these modes can grow, may differ from acceleration in electron-proton plasmas. \\

\acknowledgements We would like to thank Anatoly Spitkovsky, Uri Keshet and Avi Loeb for useful discussions.
This research was partially supported by ISF, AEC and Minerva grants.

\appendix

\section{Full solution of the dispersion equation}
\label{sec:DispEqSol}

Consider the case, described in \sref{sec:analysis}, where the velocity distribution of the particles is axi-symmetric (around $x$).
In this case averages over the velocity distributions of the type $\langle\bt_{\perp}\rangle$, and $\langle\bt_z\rangle$, where $\perp,z$ are directions perpendicular to x, are zero.
As a result, terms in the beam susceptibility \citep[c.f.][]{Melrose86}, $\chi_{CR}$ containing such averages cancel out, and the susceptibility will only have non-diagonal term in the plane defined by beam direction and the wavenumber vector, $\bk$ (the $x-k$ plane).
In this plane the beam susceptibility, the electro-magnetic field susceptibility, and the cold plasma susceptibility have matrix forms which are respectively
\be
    \chi_{CR}=\ocr ^{2}\left(\begin{array}{cc}
    1 - \frac{2\beta k_{||}}{\Omega}+\frac{(k^{2}-\omega^{2})\beta^{2}}{\Omega^{2}} &
    -\frac{\beta k_{\perp}\omega^{2}}{\Omega}\\
    -\frac{\beta k_{\perp}\omega^{2}}{\Omega} & 1\end{array}\right); \quad
    \chi_{EM}=-\left(\begin{array}{cc}
    \omega^{2}-k_{\perp}^{2} & k_{||}k_{\perp}\\
    k_{||}k_{\perp} & \omega^{2}-k_{||}^{2}\end{array}\right);
    \quad
    \chi_{0}=\oo^{2},
\ee
where $\Omega$ is defined after equation \eqref{eq:Jcr_parN}, and the magnetic field $\bB$ is perpendicular to the $x-k$ axis.
In the remaining axis only the plasma frequency term remains and this axis gives rise only to the plasma oscillation modes.

In the absence of the beam the dispersion equation,
$\det{ \left\{ \chi_{0} + \chi_{EM} \right\} } = 0$, is a forth order polynomial equation in $\om$ with real coefficients  which has the following 4 solutions, $\om=\pm \oo, \pm (\oo^2 + k^2)^{0.5}$, that represent respectively the plasma oscillation and the electro-magnetic mode.
In the presence of the beam the dispersion equation becomes
\be
\label {eq:FullDisp}
\det{ \left\{ \chi_{0} + \chi_{CR} + \chi_{EM} \right\} } = 0,
\ee
which is a 6'th order polynomial equation in $\om$ with real coefficients. This equation is a slight perturbation of the original dispersion equation with 6 distinct solutions.
Four of these solutions are slight deviation from the solutions of the original dispersion equation, while the other two solutions, for the $k$ space regime that is considered above, are the solutions discussed in \sref{sec:nospread}.
The full dispersion equation can be solved numerically for any $\bk$ and these solutions are shown in figure \eqref{fig:GrDeltaFunc}.

\section{1-mode PIC simulations}
\label{sec:NumMethods}

The solution described above was numerically verified to be the fastest growing one at several wavevectors, $\bold{k}$.
This was done by performing an efficient PIC simulation in which only one $\bold{k}$ mode is treated while the rest of the modes are neglected.
In this simulation, the electric field, magnetic field, and electric current are sinusoidal with a given $\vec{k}$ value and a time dependent amplitude, while the cosmic rays are treated as particles with continuous position (1D along $\vec{k}$) and momenta (3D).
The upstream, which is assumed to have a delta function momentum distribution, is written in terms of fluid quantities in the linear approximation and is likewise sinusoidal with the given $\vec{k}$.
The time dependent amplitude of the electric current carried by the cosmic rays is derived from the distribution of the particles by
\begin{equation}
    \bold{j}^{CR}_{\bold{k}} =
    \sum_j q_j \boldsymbol{\beta}_j
    \exp{\left(\bold{k}\cdot \bold{x}_j \right)},
\end{equation}
where $ q_j, \boldsymbol{\beta}_j, \text{and }\bold{x}_j $ are the charge velocity and position of the particle $j$.
As such, the simulation is only accurate for the linear regime (the amplitude at which the mode becomes non-linear can in principle be identified).
The growth rate is calculated by fitting the time evolution of the amplitude of the mode with an exponential function in time.
With this method only the fastest growing mode at the given $\vec{k}$ is accounted for.
This is a direct, physically transparent method of studying the linear regime of a single wavevector, with no restriction on the 3D velocity distribution, which is time and memory efficient.

\begin{figure}
\includegraphics[scale=1]{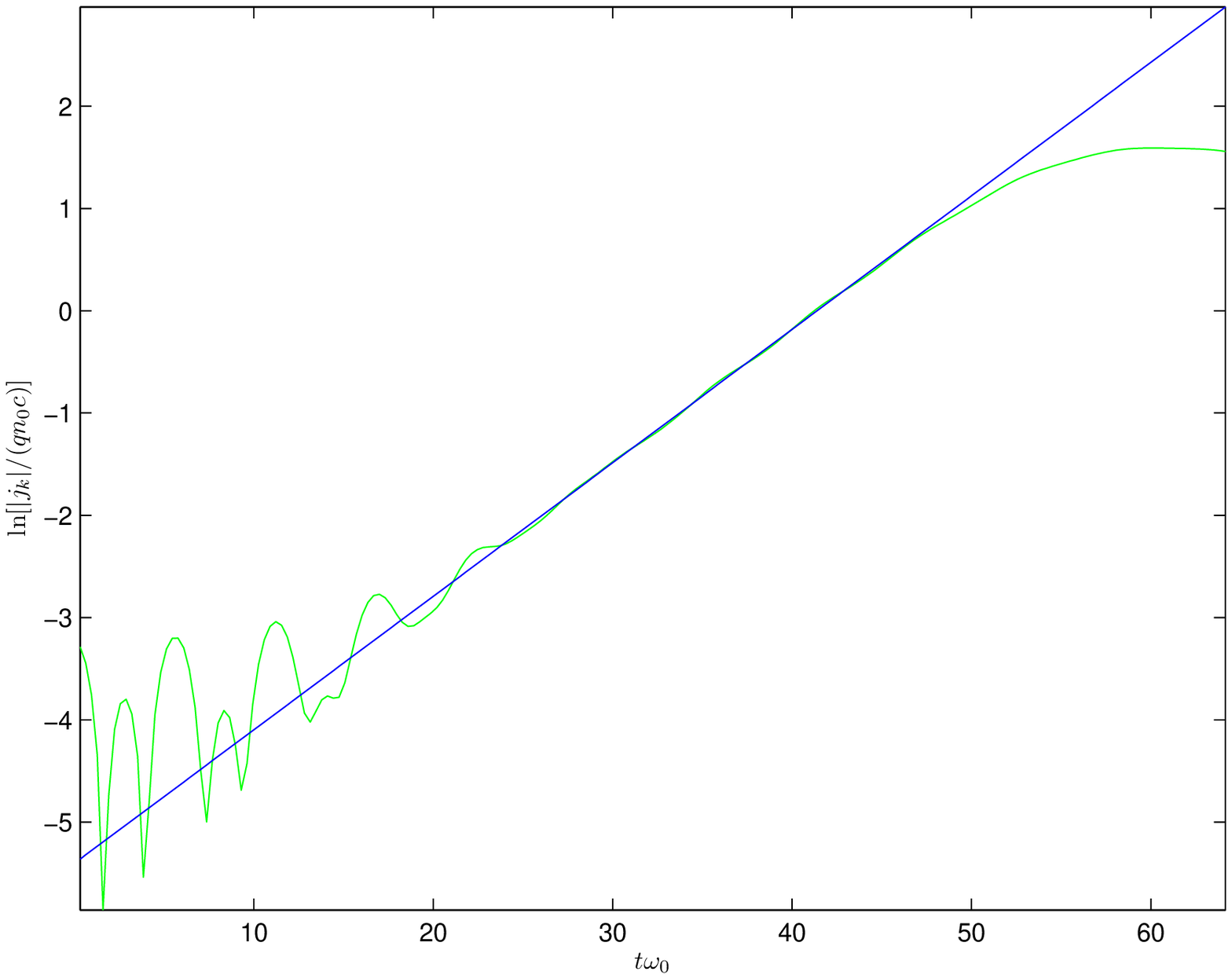}
\caption{\label{fig:AppendSimRes}
The growth of a $\{k_{||}/\oo=0.05,k_{\perp}/\oo = 0.5\}$ Weible mode in a 1-mode PIC simulation described in the text in \sref{sec:NumMethods}.
The parameters chosen for this simulation were taken to reproduce the $\gamma_{s} = 25$ ($\ocr^2 = 0.1, \gamma_u = 1250$) point given in figure \eqref{fig:GRforRealBeam}.
The green curve is the logarithm of the amplitude of the current vector, $j_k$, obtained from the simulation, and the blue line is a linear fit to this curve in the range
$19.6 < t\oo < 44.5$.
Its slope, $ 0.131$, represents the growth rate.
}

\end{figure}

For illustration, the results of the simulation used for figure \eqref{fig:GRforRealBeam} are presented in figure \eqref{fig:AppendSimRes}.
As an underlaying quantity for the fitting we use the amplitude of the electric current, $j_k$, which is normalized to the current $j_0 = q n_0 c$, where $n_0$ is the upstream density, and $q$ is the electron charge.
As can be seen in the figure, the current $j_k$ grows exponentially with time and the growth rate of the instability is easily obtained from the fit.
The accuracy of the growth rate obtained numerically is high despite the fact that only $10^6$ particles (cosmic rays) were used.
The saturation level of $j_k$ can also be obtained from this figure. However, since all quantities, except those related to the cosmic rays, are treated in the linear regime, this saturation is representative only if in reality the saturation is governed by the cosmic rays.

\bibliographystyle{apj}
\bibliography{general}

\end{document}